\documentclass[twocolumn]{aastex61}
\pdfoutput=1
\usepackage{amsmath}
\usepackage{url}

\newcommand{\msun} {M_\odot}
\newcommand{\lsun} {L_\odot}
\newcommand{\Teff} {T_{\rm eff}}
\newcommand\gta{\lower 0.5ex\hbox{$\ \buildrel > \over \sim\ $}}
\newcommand\lta{\lower 0.5ex\hbox{$\ \buildrel < \over \sim\ $}}

\begin{document}

\title{On the Spectral Evolution of Hot White Dwarf Stars. \\ IV. The Diffusion and Mixing of Residual Hydrogen in Helium-rich White Dwarfs}

\author{A. B\'edard}
\affiliation{Department of Physics, University of Warwick, Coventry, CV4 7AL, UK; email: antoine.bedard@warwick.ac.uk}
\affiliation{D\'epartement de Physique, Universit\'e de Montr\'eal, Montr\'eal, QC H3C 3J7, Canada \vspace*{6mm}}

\author{P. Bergeron}
\affiliation{D\'epartement de Physique, Universit\'e de Montr\'eal, Montr\'eal, QC H3C 3J7, Canada \vspace*{6mm}}

\author{P. Brassard}
\affiliation{D\'epartement de Physique, Universit\'e de Montr\'eal, Montr\'eal, QC H3C 3J7, Canada \vspace*{6mm}}

\shorttitle{Hydrogen in Helium-rich White Dwarfs}
\shortauthors{B\'edard, Bergeron \& Brassard}

\begin{abstract}

In the framework of our extensive modeling study of the spectral evolution of white dwarfs, we present here a new set of detailed calculations of the transport of residual hydrogen in helium-rich white dwarfs. First, we investigate the so-called float-up process at high effective temperature, whereby the upward diffusion of trace hydrogen leads to the formation of a hydrogen atmosphere. We examine the dependence of this phenomenon on the initial hydrogen abundance and on the strength of the radiative wind that opposes gravitational settling. Combined with our empirical knowledge of spectral evolution, our simulations provide new quantitative constraints on the hydrogen content of the hot helium-dominated white dwarf population. Then, we study the outcome of the so-called convective dilution process at low effective temperature, whereby the superficial hydrogen layer is mixed within the underlying helium-rich envelope. In stark contrast with previous works on convective dilution, we demonstrate that, under reasonable assumptions, our models successfully reproduce the observed atmospheric composition of cool DBA stars, thereby solving one of the most important problems of spectral evolution theory. This major improvement is due to our self-consistent modeling of the earlier float-up process, which predicts the existence of a massive hydrogen reservoir underneath the thin superficial layer. We argue that the trace hydrogen detected at the surface of DBA white dwarfs is, in most cases, of primordial origin rather than the result of external accretion.

\end{abstract}

\section{Introduction} \label{sec:intro}

The evolution of white dwarf stars is often viewed as a simple phenomenon, given that these stellar remnants are almost devoid of energy sources and thereby continuously cool with time. Moreover, as a result of their extreme compactness, the process of gravitational settling effectively removes all heavy elements from their outer layers, which are thus usually characterized by a high degree of chemical purity. Nevertheless, a closer look at these objects reveals many complexities, and in particular a surprisingly rich variety of surface compositions. Besides the standard atmospheres made of pure hydrogen (type DA\footnote{Although hydrogen-atmosphere white dwarfs are generally of type DA, the coolest of them are actually of type DC given that hydrogen lines disappear below $\Teff \sim 5000$ K.}), atmospheres dominated by helium (types DO, DB, DC) and/or contaminated by trace elements (types DBA, DQ, DZ, and many others) are also quite common among the white dwarf population. Even more interestingly, the nature of the main surface constituent can change with time, from helium to hydrogen and from hydrogen to helium. This so-called spectral evolution is thought to arise from a collection of chemical transport mechanisms, such as atomic diffusion, convective mixing, stellar winds, and matter accretion, which can compete against gravitational settling in different phases of the life of a white dwarf. The study of this intricate phenomenon is fascinating in its own right, but is also of key importance for the use of white dwarfs as cosmochronometers, given that the envelope composition influences the cooling rate. It has been the purpose of the present series of papers to improve our understanding of spectral evolution, both empirically (\citealt{bedard2020}, hereafter Paper I) and theoretically (\citealt{bedard2022a,bedard2022b}, hereafter Papers II and III).

From an empirical point of view, the most straightforward way to investigate the atmospheric transformations is to track the fraction of helium-rich white dwarfs as a function of effective temperature. This has been done repeatedly over the years using increasingly sophisticated datasets, models, and techniques (\citealt{sion1984,greenstein1986,fleming1986,fontaine1987,dreizler1996,bergeron1997,bergeron2001,bergeron2011,napiwotzki1999,eisenstein2006,tremblay2008,krzesinski2009,giammichele2012,limoges2015,ourique2019,blouin2019,genest-beaulieu2019,cunningham2020,mccleery2020}; Paper I; \citealt{lopez-sanjuan2022}). At high temperature, we have shown in Paper I that the fraction of helium-atmosphere objects gradually decreases from $\sim$24\% at $\Teff \gta 70,000$ K to $\sim$8\% at $\Teff \sim 30,000$ K. At low temperature, the most recent analyses have established that this proportion remains roughly constant down to $\Teff \sim 20,000$ K and then rises again among cooler stars. Although there are mild differences from one study to another, the general view is that the incidence of helium-dominated atmospheres reaches $\sim$20$-$40\% at the end of the cooling sequence (see Figure 9 of \citealt{lopez-sanjuan2022} for a compilation of modern results). Taken together, these numbers suggest that (1) $\sim$15$-$30\% of white dwarfs initially have a helium-rich atmosphere but experience a helium-to-hydrogen transition at high temperature and a hydrogen-to-helium transition at low temperature, and that (2) $\sim$5$-$10\% of white dwarfs retain a helium-rich surface throughout their entire evolution. As for the remaining objects, they likely always possess a standard hydrogen-dominated atmosphere.

The commonly accepted physical interpretation of spectral evolution dates back, in essence, to \citet{fontaine1987}. In their scenario, the transformation of a hot helium-atmosphere DO/DB white dwarf into a hydrogen-atmosphere DA white dwarf is the consequence of the so-called float-up process. The basic assumption is that a small amount of residual hydrogen is initially diluted (and thus hidden) within the hot helium-rich envelope. As the star cools, this hydrogen gradually diffuses upward under the influence of gravitational settling and eventually forms a thin pure-hydrogen layer at the surface. At lower temperature, the conversion of this hydrogen-atmosphere DA white dwarf into a helium-atmosphere DB/DC white dwarf is directly linked to the development of a convection zone in the envelope. Two similar but distinct mechanisms are usually invoked. On one hand, the so-called convective dilution process is believed to take place if the mass of the superficial hydrogen shell is smaller than $\sim$10$^{-14} M$, where $M$ is the total mass of the star \citep{rolland2018}. In this case, the underlying helium mantle becomes convectively unstable; the convective motions then erode the hydrogen layer from below until the latter is completely diluted within the helium-rich envelope. On the other hand, a superficial hydrogen layer thicker than $\sim$10$^{-14} M$ gives rise to the so-called convective mixing process \citep{rolland2018}. In this case, it is in the hydrogen layer, rather than in the helium shell underneath, that a convection zone forms. The convective region deepens with cooling and ultimately reaches the helium mantle, which results in a thorough mixing of the hydrogen and helium layers. Because the helium reservoir is always orders of magnitude more massive, both mechanisms produce a helium-dominated surface with merely a trace of hydrogen. Convective dilution and convective mixing occur over different but complementary temperature ranges ($30,000 \ {\rm K} \gta \Teff \gta 14,000$ K and $14,000 \ {\rm K} \gta \Teff \gta 6000$ K, respectively; \citealt{rolland2018,rolland2020,cunningham2020}), so it is thought that both phenomena contribute to the observed spectral evolution.

From these considerations, it is clear that the spectral evolution of a white dwarf depends primarily on its hydrogen content. Canonical models of stellar evolution predict that the total mass of hydrogen left on the cooling sequence should be $\sim$10$^{-4} M$ \citep{iben1984,renedo2010}. Such a ``thick'' hydrogen layer precludes any of the atmospheric transformations described above and is thus probably representative of those standard white dwarfs which perpetually exhibit a hydrogen-rich surface. In contrast, the stars entering the cooling sequence with a helium-rich atmosphere must have lost nearly all of their hydrogen in previous evolutionary phases, most likely through either a late helium-shell flash \citep{iben1983,herwig1999,althaus2005a,werner2006} or a stellar merger \citep{zhang2012,reindl2014a}. However, the majority of these objects must retain at least some hydrogen, given that they later develop a superficial hydrogen layer. The amount of hydrogen left in the star largely determines when (or alternatively, at what effective temperature) the spectral changes take place. At high temperature, the more hydrogen is contained within the helium-rich envelope, the earlier the float-up process leads to a DO/DB-to-DA transition. At low temperature, the thicker the superficial hydrogen layer, the later convective dilution or mixing occurs and produces a DA-to-DB/DC transition. Furthermore, the amount of hydrogen present in the star also dictates the hydrogen abundance in the helium-rich envelope after convective dilution or mixing has happened. Therefore, the total mass of residual hydrogen in helium-dominated white dwarfs is a central question in the study of spectral evolution \citep{fontaine1987,macdonald1991,tremblay2008,bergeron2011,koester2015,rolland2018,rolland2020,cunningham2020}.

Although the empirical analyses mentioned above provide useful information on the incidence and time of the spectral transformations, they do not directly constrain the total mass of hydrogen because they only probe the composition of the visible atmosphere. To hope to obtain such constraints, empirical findings must be combined with theoretical modeling of the whole stellar envelope, ideally taking into account both stellar evolution and element transport in a self-consistent way\footnote{We note that asteroseismology can also provide independent constraints on the total mass of hydrogen in individual pulsating white dwarfs \citep{giammichele2016,corsico2019}.}. Nevertheless, this is notoriously challenging from a numerical point of view, mainly because of the huge disparity between the cooling and transport timescales of white dwarfs. For this reason, most theoretical studies of spectral evolution published so far have made some approximations in the treatment of chemical transport.

The float-up scenario at high temperature was investigated by \citet{macdonald1991}, who however did not simulate the float-up process per se. They instead relied on sequences of static envelope models in diffusive equilibrium, where all of the hydrogen is assumed to have already reached the surface. This assumption was later shown to be incorrect: although gravitational settling is highly efficient in the very outer layers, it operates on a much longer timescale at the bottom of the envelope, where hydrogen can thus remain hidden for a long period of time \citep{dehner1995,fontaine2002,althaus2004}. For this reason, the study of \citet{macdonald1991} only provides a crude lower limit on the mass of hydrogen required for the DO/DB-to-DA transition to take place ($\mkern-5mu \gta$10$^{-15} M$). The first detailed calculations of the diffusion of hydrogen in hot helium-rich white dwarfs were performed by \citet{unglaub1998,unglaub2000}. These authors still used static envelope models but solved the full time-dependent transport problem. They demonstrated that the DO/DB-to-DA transition indeed occurs extremely rapidly under the influence of gravitational settling alone, but can be significantly delayed by mass loss due to a weak radiative wind. As such, they showed that the float-up phenomenon is primarily governed by three parameters: the initial hydrogen abundance, the wind mass-loss rate, and the stellar mass. Nevertheless, they explored a limited region of the parameter space, as they only considered a few (relatively large) hydrogen abundance values and did not vary the mass-loss law. For instance, they found that at $M = 0.6 \ \msun$, hot DO white dwarfs with initial hydrogen mass fractions $X_{\rm H,0} \sim 10^{-2}$ and $10^{-3}$ (where the index 0 refers to values at the beginning of the cooling sequence) turn into DA white dwarfs at $\Teff \sim 85,000$ and 70,000 K, respectively\footnote{While in this paper we generally express the hydrogen abundance as a mass fraction (simply because this is the input quantity of our calculations), \citet{unglaub1998,unglaub2000} instead use number fractions. In the text, we refer to their calculations assuming $N_{\rm H,0}/N_{\rm He,0} = 10^{-1}$ and $10^{-2}$, which approximately correspond to $X_{\rm H,0} \sim 10^{-2}$ and $10^{-3}$, respectively, given their adopted composition (see Section 7 of \citealt{unglaub2000}).}. Since the work of \citet{unglaub1998,unglaub2000}, very few other theoretical investigations of the kind have been conducted. In Paper II, we presented full evolutionary calculations including a self-consistent, time-dependent treatment of element transport, with the specific aim of modeling the spectral evolution of white dwarfs. In particular, we analyzed in detail a single simulation of the float-up process and the associated DO/DB-to-DA transition, leaving a thorough examination of the parameter space for future work. Similar computations had also been carried out previously by \citet{althaus2005b,althaus2020}, but these authors discussed only briefly the unfolding of the chemical transformation itself, focusing instead on the final outcome and its implications for stellar pulsations. All in all, the study of the float-up of residual hydrogen has remained fragmentary at best, and therefore its potential for probing the hydrogen content of hot helium-rich white dwarfs is still largely unexploited.

On the other hand, the convective dilution and convective mixing mechanisms at low temperature have been the topic of several theoretical papers (\citealt{baglin1973,koester1976,dantona1989,macdonald1991,althaus1998,chen2011,rolland2018,rolland2020}; Paper II). The most puzzling result of these works is the long-standing problem of the origin of hydrogen in cool DBA stars. These helium-dominated, hydrogen-bearing objects are mostly found in the range $20,000 \ {\rm K} \gta \Teff \gta 12,000$ K (where they account for the majority of the helium-rich white dwarf population) and typically have surface hydrogen abundances (by number) in the range $-6.5 \lta \log N_{\rm H}/N_{\rm He} \lta -4.0$ \citep{beauchamp1996,voss2007,bergeron2011,koester2015,rolland2018,genest-beaulieu2019}. They are commonly believed to be the outcome of the convective dilution process. Yet, traditional models of this phenomenon predict final hydrogen abundances that are orders of magnitude lower than those measured in DBA white dwarfs. Consequently, an additional source of hydrogen (that is, besides the thin pure-hydrogen layer once present at the surface) must be invoked to explain these observations. It has often been proposed that the hydrogen is externally accreted, either from the interstellar medium \citep{macdonald1991,voss2007} or from water-rich comets, asteroids, or planets \citep{farihi2013,veras2014,raddi2015,gentile-fusillo2017}. In recent years, the second hypothesis has been shown to successfully account for the composition of some DBA white dwarfs, especially those with very large hydrogen abundances ($\log N_{\rm H}/N_{\rm He} \gta -4.0$). Nevertheless, the accretion scenario faces a difficulty when applied to the bulk of the DBA population: it requires that accretion begins only after convective dilution has happened. Otherwise, the superficial hydrogen layer built up by accretion rapidly becomes too thick to be altered by convection, and thus the star retains a hydrogen-dominated atmosphere \citep{bergeron2011,koester2015,rolland2018}. Meanwhile, \citet{rolland2020} put forward a completely different paradigm: they suggested that the source of hydrogen is internal rather than external. More specifically, they argued that the residual hydrogen located at great depths, where the diffusion timescales are very long, does not have enough time to float to the surface. Once convective dilution has taken place, this reservoir of hydrogen finds itself mixed within the convection zone and thereby affects the final atmospheric composition\footnote{\citet{rolland2020} refer to this phenomenon as a dredge-up process, but we argue in Section \ref{sec:res_conv} that this term is somewhat of a misnomer as we really are dealing with a dilution process.}. \citet{rolland2020} tested this idea quantitatively and demonstrated that it is promising. However, they relied on static envelope models with approximate chemical profiles, so a more detailed theoretical investigation is definitely warranted.

In this paper, we study the transport of residual hydrogen in hot helium-rich white dwarfs with the aim of constraining the hydrogen content of these objects. To do so, we carry out state-of-the-art simulations of the float-up process analogous to that introduced in Paper II. We examine, in particular, how the initial hydrogen abundance influences the DO/DB-to-DA transition. We can then estimate the range of hydrogen content that characterizes the helium-dominated white dwarf population based on the empirical results of Paper I. Furthermore, a secondary purpose of the present work is to revisit the problem of the surface composition of cool DBA white dwarfs using our improved evolutionary models. More specifically, we investigate the convective dilution scenario proposed by \citet{rolland2020} using the realistic chemical profiles obtained from our float-up simulations. In Section \ref{sec:comp}, we describe the physical ingredients of our computations. Then, our results are presented in Section \ref{sec:res} and discussed in Section \ref{sec:disc}. Finally, our conclusions are summarized in Section \ref{sec:conclu}.

\section{Computations} \label{sec:comp}

As in the previous papers of this series, we use the STELUM evolutionary code to perform our spectral evolution simulations. A thorough description of the constitutive physics and numerical techniques of the code can be found in Paper II. Briefly, we build complete white dwarf models, from the center to the surface, and evolve them over time while allowing the chemical structure to change. The transport of chemical elements is modeled using a time-dependent diffusive approach and is self-consistently coupled to the evolution of the star, meaning that the feedback of composition changes on the cooling process is taken into account. The transport mechanisms included in our calculations are atomic diffusion, large-scale mixing due to convection, and a weak radiative wind.

Our starting models have a mass $M = 0.6 \ \msun$, an effective temperature $\Teff = 100,000$ K, a homogeneous carbon/oxygen core, and a homogeneous hydrogen-deficient envelope. We study two families of models characterized by different initial envelope compositions: a largely helium-dominated plasma and a helium/carbon/oxygen mixture. These two compositions are representative of the two classes of hydrogen-deficient white dwarf progenitors, the O(He) stars and the PG 1159 stars, respectively \citep{werner2006,reindl2014a}. In both cases, we also include a uniform trace of hydrogen in the envelope. The initial mass fraction of hydrogen $X_{\rm H,0}$ is an input parameter and the values considered in this work span the range $-7.0 \le \log X_{\rm H,0} \le -2.0$ (in steps of 0.5 dex). The initial mass fractions of carbon and oxygen are $X_{\rm C,0} = X_{\rm O,0} = 0.001$ for the O(He)-type models, and $X_{\rm C,0} = 0.5$ and $X_{\rm O,0} = 0.1$ for the PG 1159-type models. The remaining material is helium ($X_{\rm He,0} = 1 - X_{\rm H,0} - X_{\rm C,0} - X_{\rm O,0}$). The exact proportions of carbon and oxygen are inconsequential, given that these elements quickly sink into the star and leave behind a hydrogen/helium mixture; nevertheless, our two types of composition are expected to power stellar winds of different strengths, an effect that we take into consideration below. In terms of the fractional mass depth $q = 1-m/M$ (where $m$ is the usual interior mass), the initially homogeneous envelope extends down to $\log q = -2.0$, with the exception that we include residual hydrogen only above $\log q = -4.0$, because deeper-lying hydrogen is burned in pre-white dwarf evolutionary phases \citep{althaus2005b,renedo2010}.

The float-up process is essentially governed by an interplay between atomic diffusion, which carries hydrogen upward, and the radiative wind, which tends to maintain a homogeneous composition in the outer layers. Thus, these two transport mechanisms must be treated as realistically as possible. Our simulations include the three types of particle diffusion (chemical diffusion, gravitational settling, and thermal diffusion) following the formalism of \citet{burgers1969} and using the diffusion coefficients of \citet{fontaine2015}. In particular, we have shown in Paper III (see also \citealt{althaus2004}) that thermal diffusion, which is often considered negligible in white dwarfs, has a significant impact on spectral evolution at high temperature. Unlike in our previous papers, we ignore here the non-ideal term of \citet{beznogov2013} in the diffusion equations, as it becomes important only at low temperature.

Ideally, the wind would be handled through the theory of radiation-driven winds, but this approach is highly impractical in an evolutionary context. One must then resort to approximate analytic formulas to evaluate the wind mass-loss rate efficiently along an evolutionary sequence. In the case of hot white dwarfs, a few studies have demonstrated that the simple mass-loss law of \citet{blocker1995},
\begin{equation}
\dot{M} = -1.29 \times 10^{-15} \left( \frac{L}{\lsun} \right)^{1.86} \msun \ \rm{yr}^{-1} ,
\label{eq:mdot}
\end{equation}
where $L$ is the surface luminosity, accounts relatively well for the observed properties of PG 1159 stars and their descendants (\citealt{unglaub2000,quirion2012}; Paper III). Therefore, we adopt this expression for our PG 1159-type models, and we henceforth refer to it as the strong wind or the high mass-loss rate. As for objects with almost pure-helium envelopes, the situation is even less clear, because there exists no definite constraint on the strength of their winds. The best we can do is to use Equation \ref{eq:mdot} scaled by a numerical factor chosen according to our knowledge of the winds of their direct precursors, the O(He) stars. Based on a rough comparison between the estimated mass-loss rates of individual O(He) \citep{reindl2014a} and PG 1159 \citep{koesterke1998,herald2005} pre-white dwarfs, we simply set the mass-loss rate of our helium-dominated models to one-tenth the value given by Equation \ref{eq:mdot}. This is admittedly nothing more than an educated guess, but this allows us to investigate the effect of the wind prescription on the float-up of hydrogen. In the following, this reduced version of the mass-loss law is called the weak wind or the low mass-loss rate. Note that in either case, diffusion is severely inhibited by the wind at $\Teff \gta 100,000$ K, so the initial effective temperature adopted in our simulations is high enough to capture the entirety of the float-up process.

\begin{figure}
\centering
\includegraphics[width=\columnwidth,clip=true,trim=5.75cm 8.00cm 5.75cm 10.00cm]{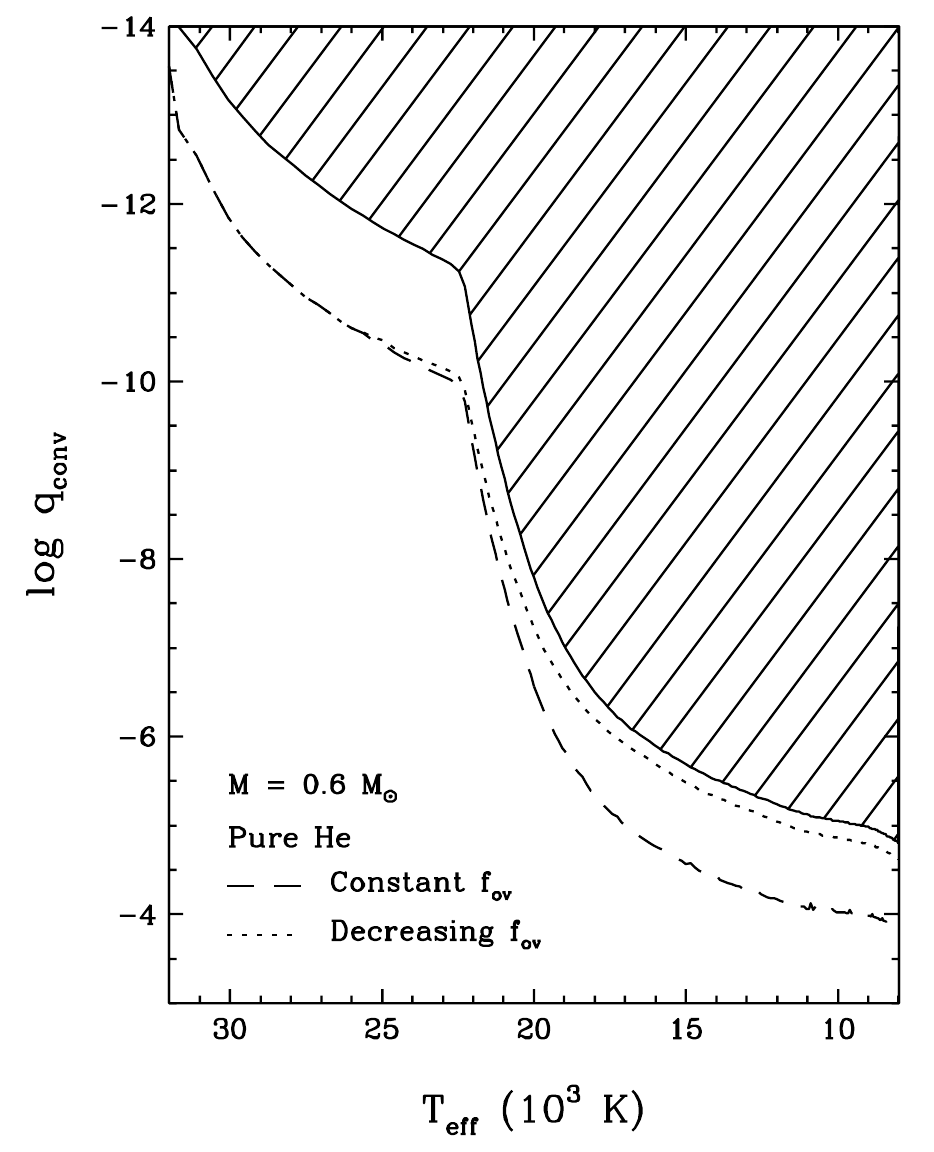}
\caption{Extent of the convectively mixed region as a function of effective temperature in a white dwarf model with a mass $M = 0.6 \ \msun$ and a pure-helium envelope. The position inside the star is measured in terms of the fractional mass depth ($q = 1-m/M$). The hatched area represents the formally convective zone, while the dashed and dotted curves show the base of the overshoot region for a constant and decreasing overshoot parameter, respectively (see text for details).}
\vspace{2mm}
\label{fig:convzone}
\end{figure}

Furthermore, we also extend some of our calculations to lower effective temperatures, where the chemical evolution is dominated by the helium convection zone and thus depends on the treatment of convective mixing. We rely on the so-called ML2 version of the mixing-length theory, in which the mixing-length parameter is $\alpha = 1.0$ \citep{bohm-vitense1958,tassoul1990}. For our purpose, which is to study cool DBA stars, the details of the mixing-length model are somewhat inconsequential because convection becomes adiabatic below $\Teff \sim 20,000$ K \citep{rolland2018,cukanovaite2019}. Much more important is the prescription for mixing due to overshoot beyond the formally convective region. As in our previous works, we use the overshoot coefficient of \citet{freytag1996},
\begin{equation}
D_{\rm ov} = D_0 \ \exp{ \left( \frac{-2 |r-r_0|}{f_{\rm ov} \, H_0} \right) } ,
\label{eq:dov}
\end{equation}
where $r$ is the radial coordinate, $r_0$ is the radius of the convective boundary, $D_0$ and $H_0$ are the mixing coefficient and pressure scale height at $r_0$, and $f_{\rm ov}$ is a numerical parameter. In this framework, it is the value of $f_{\rm ov}$ that controls the extent of the overshoot region. Unfortunately, this parameter constitutes another rather poorly constrained physical ingredient in our models. According to hydrodynamical simulations of convective overshoot in hydrogen-rich white dwarfs, the depth of the overshoot region in these objects is relatively well represented by $f_{\rm ov} \sim 0.4$ (\citealt{kupka2018,cunningham2019}; T. Cunningham 2021, private communication). Analogous calculations for helium-rich white dwarfs are not yet available but are not expected to be significantly different. Consequently, we adopt $f_{\rm ov} = 0.4$ as our default value. Nevertheless, it is possible that this value is appropriate only for the thin non-adiabatic convection zones ($\log q_{\rm conv} \lta -12.0$, where $q_{\rm conv}$ is the location of the base in terms of fractional mass depth) found in relatively hot objects ($\Teff \gta 20,000$ K). Indeed, based on observations and models of carbon-polluted DQ stars, we have shown in Paper III that cool helium-dominated white dwarfs ($\Teff \lta 10,000$ K), which have much thicker convection zones ($\log q_{\rm conv} \sim -5.0$), must have $f_{\rm ov} \le 0.075$. We interpreted this result as evidence that the relative extent of the overshoot region decreases as the convection zone deepens and becomes increasingly adiabatic. To account for this possibility, we also carry out calculations in which the overshoot parameter varies with the depth of the formally convective region: we assume $f_{\rm ov} = 0.4$ for $\log q_{\rm conv} \le -12.0$, $f_{\rm ov} = 0.05$ for $\log q_{\rm conv} \ge -6.0$, and a linear interpolation in between. For reference, Figure \ref{fig:convzone} illustrates the extent of the convectively mixed region as a function of effective temperature in a pure-helium envelope for both treatments of the overshoot parameter. Notice that with our decreasing-$f_{\rm ov}$ prescription, the mixed region is barely larger than the formally convective zone at low temperature, hence our results would not change much if we disregarded overshoot altogether.

Our simulations do not include radiative levitation, which is negligible in a hydrogen/helium plasma and therefore plays no significant role in the problem at hand \citep{vennes1988,macdonald1991}. For simplicity, we also ignore residual nuclear burning of hydrogen at the bottom of the envelope. This choice, along with our assumption on the depth of the reservoir of trace hydrogen, affects the total amount of hydrogen present in our evolving white dwarf models and thus deserves a few comments. By placing the base of the hydrogen reservoir at $\log q = -4.0$ and ignoring nuclear burning, we are effectively maximizing and even possibly overestimating the total mass of hydrogen contained within the star for a given initial hydrogen abundance. Although $\log q = -4.0$ is the standard value for the largest depth to which hydrogen can exist in typical white dwarfs \citep{renedo2010}, it is not clear that this value is suitable for the descendants of PG 1159 and O(He) stars due to their non-standard evolutionary history; for instance, Figure 3 of \citet{althaus2005b} indicates that the hydrogen reservoir may be somewhat shallower. Furthermore, nuclear burning should reduce slightly the total mass of hydrogen of our white dwarf models with time, because a small amount of hydrogen is expected to be carried below $\log q = -4.0$ by chemical diffusion (due to the sharp composition gradient there; see Figures \ref{fig:hprof_030_hot} to \ref{fig:hprof_060_hot}) and thus burned. However, we want to stress that these uncertainties have little influence on our study of the float-up process at high effective temperature. This is because the diffusion timescales are much shorter at the top than at the bottom of the envelope, and therefore the early evolution of the surface composition is largely decoupled from the base of the hydrogen reservoir (as will become apparent below; see Figures \ref{fig:hprof_030_hot} to \ref{fig:hprof_060_hot}). On the other hand, the assumed total mass of hydrogen does have an impact on the outcome of the convective dilution process at low effective temperature, as convection eventually mixes material from the deep envelope to the surface. To test the robustness of our conclusions, we explore below how the predicted atmospheric hydrogen abundance in the DBA phase is affected by a reduction of the size of the initial hydrogen reservoir. Besides, given that nuclear burning should preclude the presence of hydrogen in very deep layers, we artificially prevent hydrogen from diffusing below $\log q = -4.0$, thereby maintaining a strong abundance gradient there.

Finally, it is important to discuss briefly the outer boundary condition of the chemical structure. We do not simulate element transport up to the very surface (which is located at $\log q \sim -18.5$ in our models), as this would require an enormous amount of computing time. Rather, the transport calculations are only performed up to a fractional mass limit $q_{\rm lim}$, above which the composition is assumed to be uniform. As in our previous papers, we adopt $\log q_{\rm lim} = -14.0$. We currently cannot place the transport boundary higher in the envelope, because this gives rise to numerical instabilities. For this reason, at low effective temperature, we are unable to model the convective dilution mechanism per se, given that this phenomenon involves very thin hydrogen layers that cannot be resolved in our calculations. Nevertheless, we can still compute the subsequent evolution of the chemical profile under the assumption that convective dilution has occurred. This approach has been used in all other theoretical studies of convective dilution published so far \citep{macdonald1991,rolland2018,rolland2020} and is again used below.

\section{Results} \label{sec:res}

\subsection{The Float-up Process} \label{sec:res_float}

In this section, we describe the results of our simulations of the float-up process. We focus only on the evolution of the hydrogen abundance profile; see Section 3.2 of Paper II for a depiction of the full chemical structure in a similar calculation. That said, it should be mentioned that there is a small difference between the models of the present paper and that of Paper II with regard to the float-up process. Because the latter model was computed primarily for illustrative purposes, artificial mixing was included near the transport boundary ($\log q \lta -13.0$) to improve numerical convergence, which caused the atmospheric helium-to-hydrogen transition to be slightly delayed. We do not include such artificial mixing in the present work, and thus the quantitative results displayed below supersede those of Paper II.

\begin{figure*}
\centering
\includegraphics[width=2.\columnwidth,clip=true,trim=1.75cm 8.25cm 1.75cm 10.00cm]{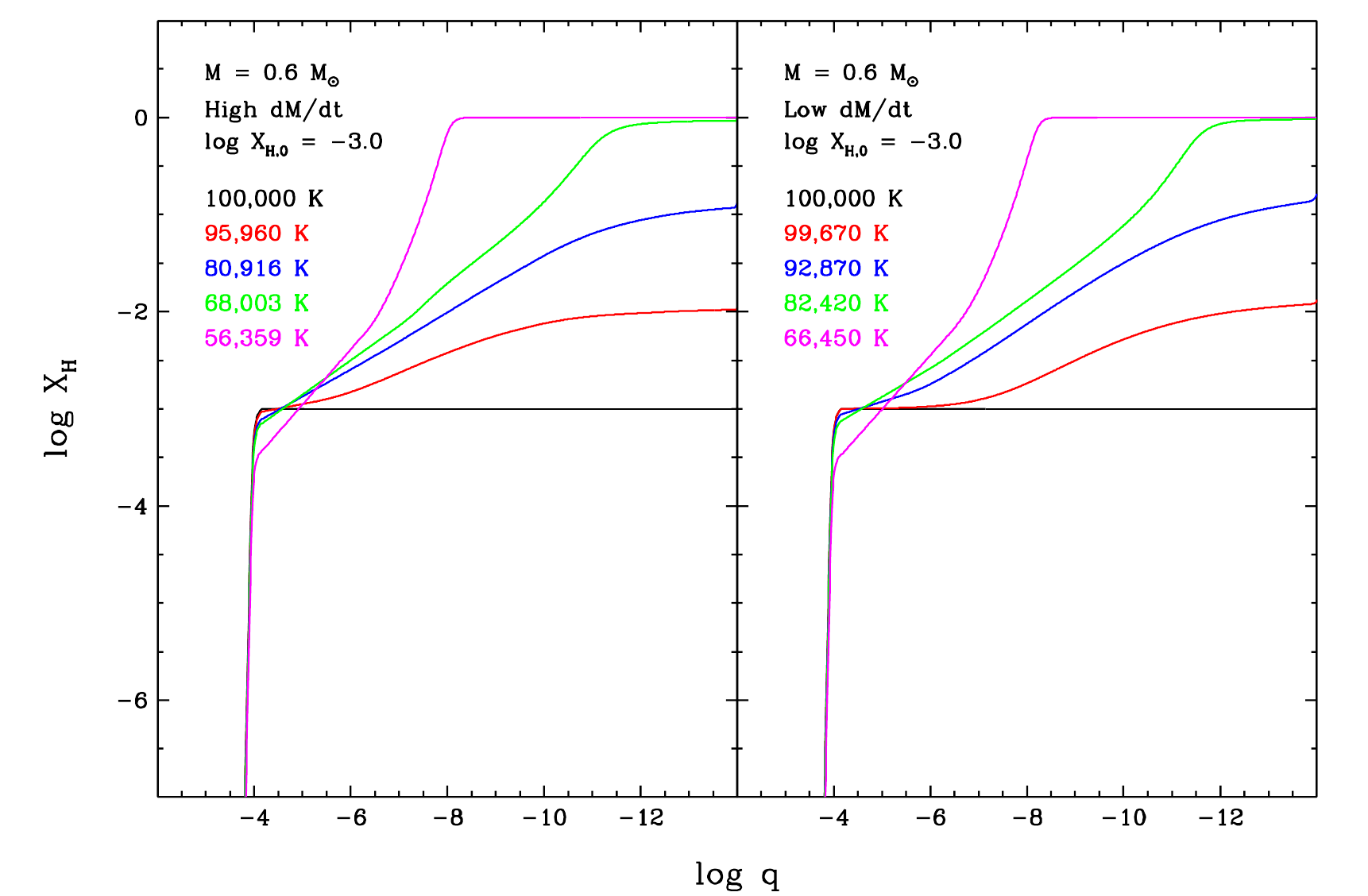}
\caption{Evolution of the hydrogen abundance profile due to the float-up process in a white dwarf model with a mass $M = 0.6 \ \msun$, a high (left panel) or low (right panel) mass-loss rate (see Section \ref{sec:comp} for details), and an initial hydrogen abundance $\log X_{\rm H,0} = -3.0$. The hydrogen abundance is expressed as a mass fraction, while the position inside the star is measured in terms of the fractional mass depth ($q = 1-m/M$). Each curve represents a selected stage along the evolutionary sequence and is labeled with the corresponding value of the effective temperature.}
\vspace{2mm}
\label{fig:hprof_030_hot}
\end{figure*}

Figure \ref{fig:hprof_030_hot} shows the evolution of the hydrogen abundance profile (the hydrogen mass fraction $X_{\rm H}$ as a function of the fractional mass depth $q$) for the initial condition $\log X_{\rm H,0} = -3.0$ and for both the PG 1159-type, strong-wind and O(He)-type, weak-wind models. Let us first concentrate on the strong-wind case, which is displayed in the left panel. At the outset of the simulation, hydrogen diffuses upward rapidly due to the highly efficient gravitational settling: the surface hydrogen abundance has already risen by $\sim$1 dex at $\Teff \sim 96,000$ K. However, it can be seen that mass loss causes the composition of the outer layers to remain nearly uniform. Consequently, as soon as a small abundance gradient is established, the float-up process is considerably slowed down by the wind. The surface hydrogen abundance has increased by another $\sim$1 dex at $\Teff \sim 80,900$ K, and the outer envelope has become hydrogen dominated at $\Teff \sim 68,000$ K. As more hydrogen from the inner envelope is carried upward by diffusion, the superficial pure-hydrogen layer thickens; for instance, it extends down to $\log q \sim -8.0$ at $\Teff \sim 56,400$ K. This specific evolutionary sequence falls within the parameter space studied by \citet{unglaub1998,unglaub2000}. The two sets of calculations are in good agreement: for the same parameters ($M = 0.6 \ \msun$, $\log X_{\rm H,0} = -3.0$, and $\dot{M}$ given by Equation \ref{eq:mdot}), they both predict that the atmosphere becomes richer in hydrogen than in helium around $\Teff \sim 70,000$ K (see Figure 19 of \citealt{unglaub2000}). In fact, the DO-to-DA transition happens slightly earlier in our model, but this is likely due the fact that we take thermal diffusion into account, while this transport mechanism was ignored by \citet{unglaub1998,unglaub2000}.

In the weak-wind case, which is shown in the right panel, the chemical evolution is qualitatively very similar but occurs much faster because gravitational settling encounters less resistance. In particular, the surface has already become hydrogen dominated at $\Teff \sim 82,400$ K, which is $\sim$14,000 K hotter than in the strong-wind case. Notice also that the lower mass-loss rate results in a slightly more pronounced abundance gradient in the outer layers of the first few models.

\begin{figure*}
\centering
\includegraphics[width=2.\columnwidth,clip=true,trim=1.75cm 8.25cm 1.75cm 10.00cm]{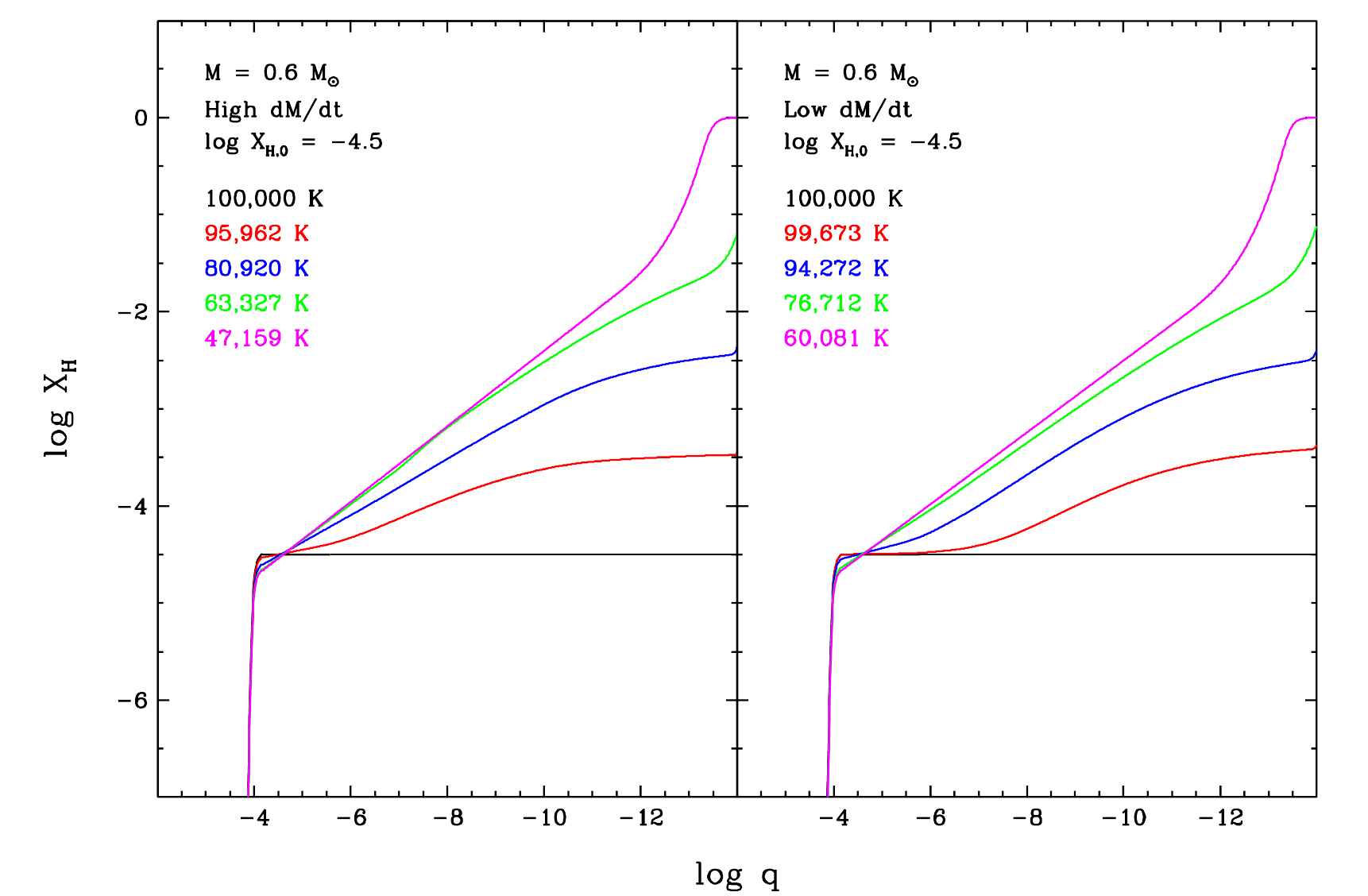}
\caption{Same as Figure \ref{fig:hprof_030_hot}, but for $\log X_{\rm H,0} = -4.5$.}
\vspace{2mm}
\label{fig:hprof_045_hot}
\end{figure*}

Figure \ref{fig:hprof_045_hot} is identical to Figure \ref{fig:hprof_030_hot}, but for $\log X_{\rm H,0} = -4.5$. All other things being equal, reducing the initial hydrogen abundance has the obvious consequence of increasing the time required for the star to develop a hydrogen-rich atmosphere. The hydrogen abundance profile initially behaves as in the previous example, but then becomes much steeper near the surface, as shown at $\Teff \sim 63,300$ and 76,700 K for the high and low mass-loss rates, respectively. This is because the wind fades with cooling and thereby looses its ability to compete against gravitational settling before the DO-to-DA transition is achieved. Hydrogen then continues to diffuse upward, such that a very thin pure-hydrogen layer has formed at $\Teff \sim 47,200$ and 60,100 K for the high and low mass-loss rates, respectively. Therefore, assuming a lower hydrogen content not only delays the DO-to-DA transition, but also allows the emergence of a much thinner superficial hydrogen layer as a result of the reduced influence of the wind in these later phases.

Another important difference with the previous example arises after the development of the outer pure-hydrogen layer: at this point, the large abundance gradient begins to drive a very small semi-convection zone around $\log q \sim -13.25$. This semi-convective region is not in a steady state: as soon as convection mixes hydrogen and helium, the composition gradient is destroyed, which causes the convection zone to disappear. Diffusion then quickly rebuilds the composition gradient, which causes the convection zone to reappear, and the cycle starts again. This oscillatory behavior occurs on an extremely short timescale (that is, less than a day); for this reason, we are numerically unable to follow the float-up process beyond this stage\footnote{In principle, STELUM has the ability to model semi-convective mixing properly (see Paper II), but the adopted scheme becomes unstable in the outer envelope, where diffusion timescales are very short.}. Nevertheless, it is likely that this semi-convective instability subsequently prevents the hydrogen located deeper in the envelope from reaching the superficial pure-hydrogen layer, which thus stops growing and remains very thin.

\begin{figure*}
\centering
\includegraphics[width=2.\columnwidth,clip=true,trim=1.75cm 8.25cm 1.75cm 10.00cm]{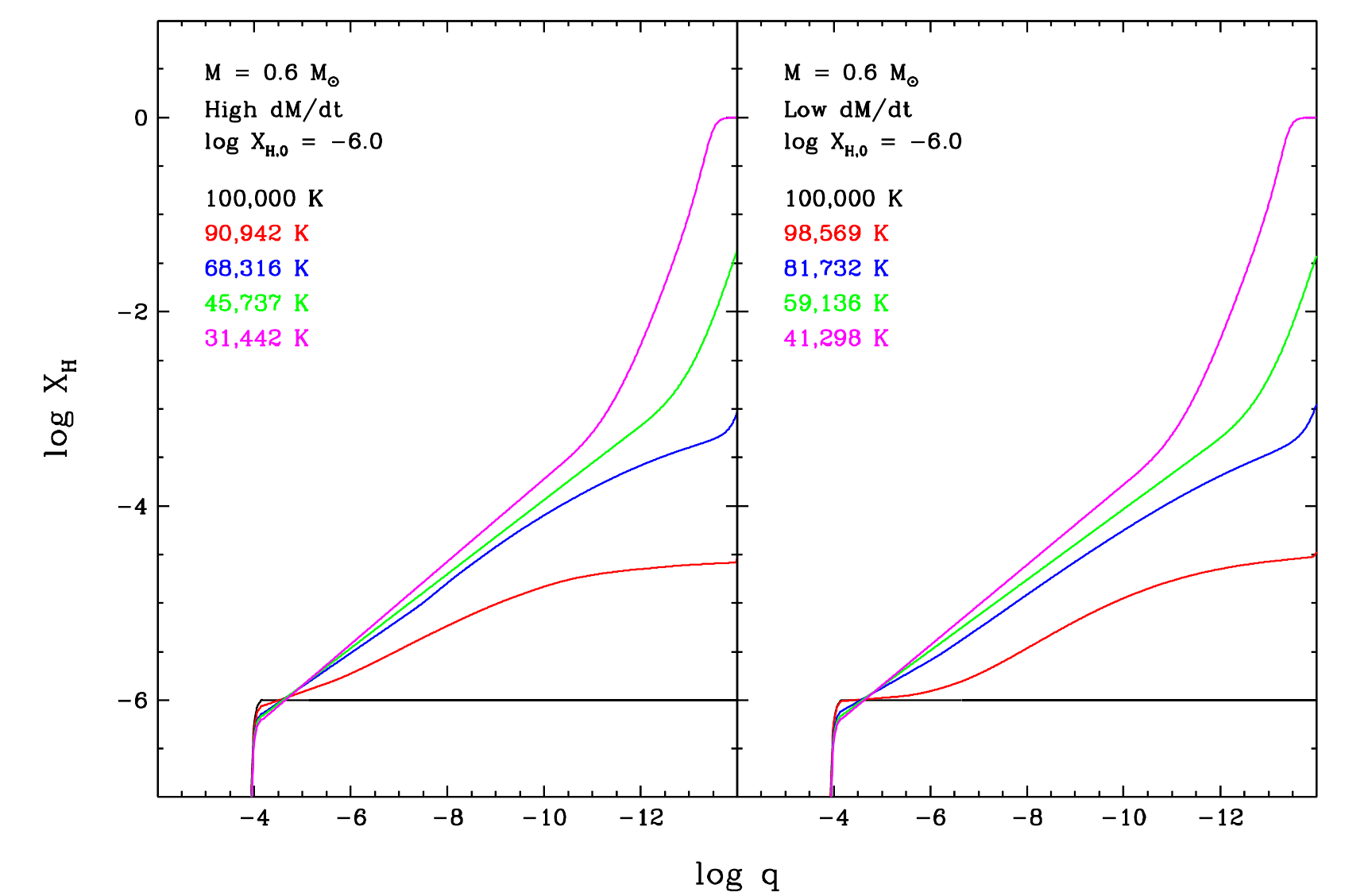}
\caption{Same as Figure \ref{fig:hprof_030_hot}, but for $\log X_{\rm H,0} = -6.0$.}
\vspace{2mm}
\label{fig:hprof_060_hot}
\end{figure*}

Figure \ref{fig:hprof_060_hot} illustrates the chemical evolution for an even lower initial hydrogen abundance, $\log X_{\rm H,0} = -6.0$. This simulation is similar to the previous one, except that the atmospheric transformation happens even later: the thin hydrogen layer emerges at $\Teff \sim 31,400$ and 41,300 K for the high and low mass-loss rates, respectively. In the strong-wind case, the float-up process is so slow that the DO star first becomes a DB star before turning into a DA white dwarf. After the DO/DB-to-DA transition, the same oscillatory semi-convective instability as before arises and precludes us from computing the subsequent chemical evolution. In fact, we find that all of our models with $\log X_{\rm H,0} \le -4.5$ experience this instability near the surface just after the formation of their hydrogen layer. Given that this phenomenon likely inhibits the upward diffusion of hydrogen, these stars should retain very thin hydrogen shells as they continue to cool. Notice that our calculations predict the existence of an extended hydrogen diffusion tail underneath the superficial layer, as envisioned by \citet{rolland2020}; this internal reservoir actually contains most of the hydrogen present in the white dwarf.

\begin{figure*}
\centering
\includegraphics[width=2.\columnwidth,clip=true,trim=1.75cm 6.75cm 1.75cm 8.75cm]{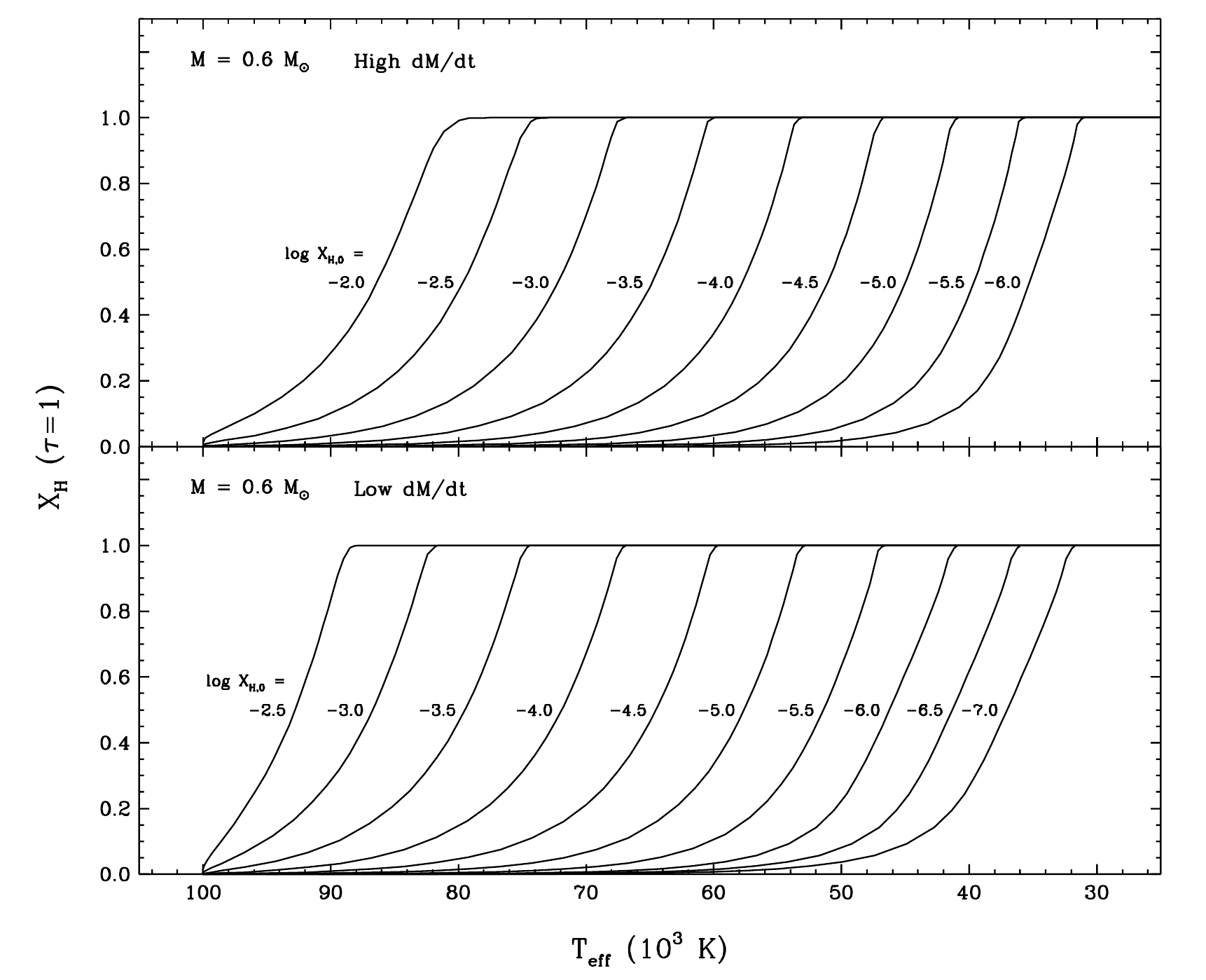}
\caption{Atmospheric hydrogen abundance as a function of effective temperature throughout the float-up process in white dwarf models with a mass $M = 0.6 \ \msun$, a high (top panel) or low (bottom panel) mass-loss rate (see Section \ref{sec:comp} for details), and various initial hydrogen abundances given in the figure. The hydrogen abundance is expressed as a mass fraction.}
\vspace{2mm}
\label{fig:evol_hot}
\end{figure*}

Figure \ref{fig:evol_hot} displays the variation of the surface hydrogen abundance with effective temperature in all of our simulations. The top and bottom panels show our PG 1159-type, strong-wind and O(He)-type, weak-wind models, respectively. This figure clearly demonstrates that the transformation of a helium-rich atmosphere into a hydrogen-rich atmosphere depends sensitively on the amount of hydrogen left in the star at the beginning of the cooling sequence. In essence, two curves differing by 0.5 dex in initial hydrogen abundance are shifted by $\sim$7000 K in effective temperature (although this shift becomes smaller for the most hydrogen-deficient models). We also note that in a given simulation, the increase in surface hydrogen abundance is quite gradual, such that there is a broad temperature range over which both hydrogen and helium are present in the atmosphere. It is therefore not surprising that white dwarfs exhibiting spectral signatures of this transitional stage have been discovered in fairly large numbers (\citealt{manseau2016}; Paper I). Furthermore, as mentioned earlier, the strength of the wind also has a marked impact on the speed of the float-up process: the curves are shifted by $\sim$10,000$-$14,000 K in effective temperature (depending on the hydrogen content) as a result of a change of a factor of 10 in the mass-loss law. Besides, note that the rightmost curve of each panel corresponds to the most hydrogen-deficient model in which the formation of a hydrogen-dominated atmosphere is possible. For lower initial hydrogen abundances, the emergence of the standard helium convection zone at $\Teff \sim 32,000$ K (see Figure \ref{fig:convzone}) halts the float-up process, hence such stars always retain a helium-dominated atmosphere.

\begin{figure}
\centering
\includegraphics[width=\columnwidth,clip=true,trim=5.75cm 8.00cm 5.75cm 10.00cm]{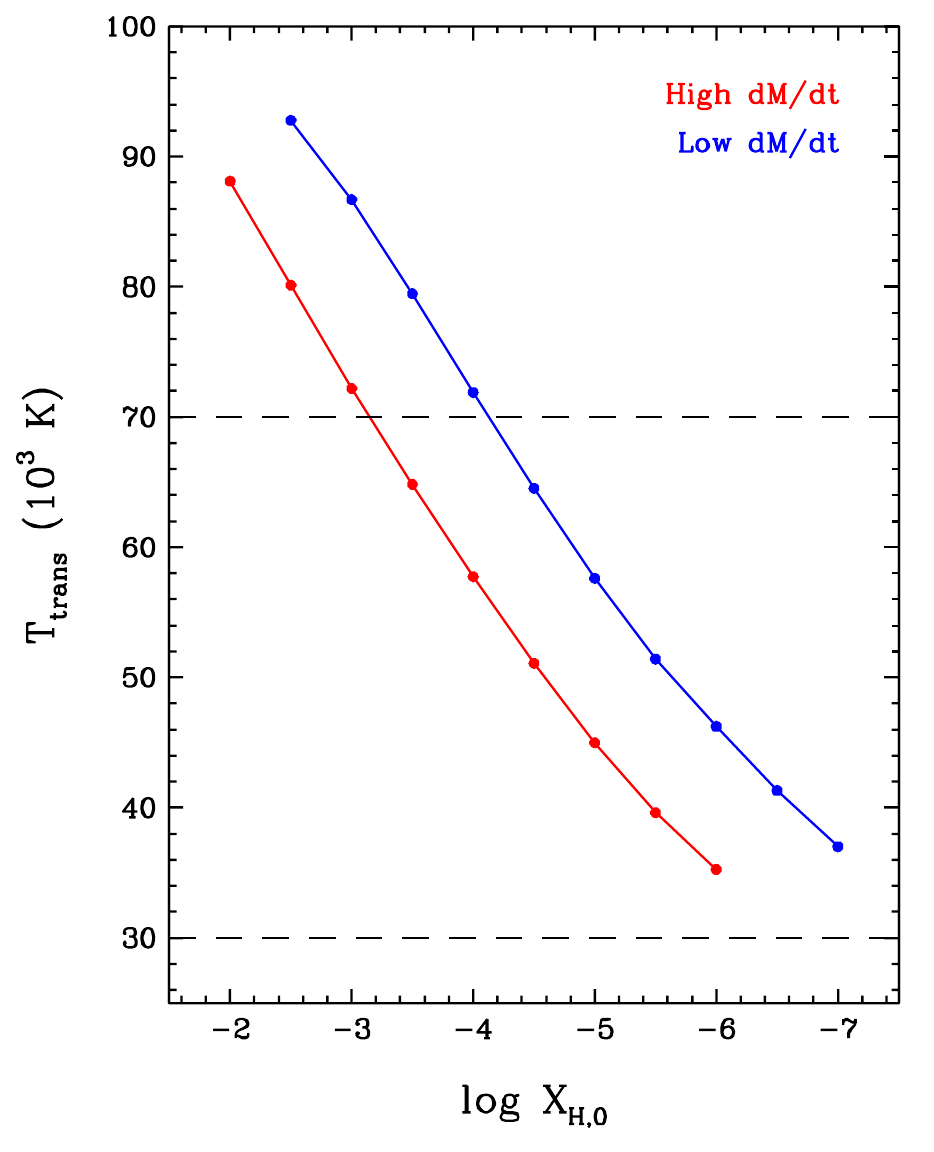}
\caption{Relation between the DO/DB-to-DA transition temperature and the initial hydrogen abundance for white dwarf models with a mass $M = 0.6 \ \msun$ and a high (red curve) or low (blue curve) mass-loss rate (see Section \ref{sec:comp} for details). The hydrogen abundance is expressed as a mass fraction. The two dashed black lines approximately delimit the range of empirical DO/DB-to-DA transition temperatures derived in Paper I.}
\vspace{2mm}
\label{fig:trans_hot}
\end{figure}

In order to examine the influence of the initial hydrogen content on the DO/DB-to-DA transition more quantitatively, let us define the transition temperature $T_{\rm trans}$ as the effective temperature for which $X_{\rm H} = X_{\rm He}$ at the surface. Figure \ref{fig:trans_hot} displays the predicted transition temperature as a function of the assumed initial hydrogen abundance for our two wind prescriptions. The two horizontal dashed lines approximately delimit the range of ``observed'' transition temperatures, that is, the range of effective temperatures over which the fraction of helium-atmosphere white dwarfs is observed to decrease continuously (Paper I). Combining our theoretical results with this empirical constraint allows us to infer the range of hydrogen content of the hot helium-rich white dwarf population. We find that the observed spectral evolution can be explained by $-6.0 \lta \log X_{\rm H,0} \lta -3.0$ if we assume a high mass-loss rate or by $-7.0 \lta \log X_{\rm H,0} \lta -4.0$ if we assume a low mass-loss rate. Accordingly, the stars that never develop a hydrogen-dominated atmosphere must have had $\log X_{\rm H,0} \lta -6.0$ or $-7.0$. Although it is clear that the treatment of the wind constitutes a major source of uncertainty, this is the first time, to our knowledge, that such constraints are obtained.

Finally, we recall that all of the calculations presented in this paper assume a fixed stellar mass $M = 0.6 \ \msun$, while it is well known that hot DO and DB stars actually span a small mass range, mostly between $M = 0.5$ and $0.7 \ \msun$ (\citealt{reindl2014b}; Paper I). The mass influences the transport of hydrogen in two ways: in a more massive white dwarf, gravitational settling is more efficient (because of the stronger gravitational field) and the radiative wind is weaker (because of the lower surface luminosity), with the result that the float-up process is faster. Thus, the fact that different objects undergo the DO/DB-to-DA transition at different effective temperatures might be partly due to differences in stellar mass (and not only to differences in hydrogen content). We carried out a few test calculations and found that varying the stellar mass by $\pm 0.1 \ \msun$ changes the transition temperature by $\pm$4000$-$6000 K (depending on the hydrogen content), which is smaller than the effect of varying the initial hydrogen abundance by $\pm 0.5$ dex. Therefore, considering a range of masses might reduce the range of hydrogen abundances required to explain the observed spectral evolution, but this effect is expected to be small.

\subsection{The Convective Dilution Process} \label{sec:res_conv}

In this section, we present evolutionary calculations at lower effective temperature with the aim of studying the implications of our float-up simulations for our understanding of cool DBA white dwarfs. These objects have long been thought to be the products of the convective dilution process in DA white dwarfs with extremely thin hydrogen layers, yet previous models of this phenomenon have failed to reproduce their observed atmospheric composition. As stated in Section \ref{sec:comp}, we cannot model the convective dilution mechanism per se within our current theoretical framework. Indeed, our choice of transport boundary ($\log q_{\rm lim} = -14.0$) prevents us from resolving the very outer parts of the envelope where convective dilution is expected to take place. Nevertheless, one way around this problem is to assume that convective dilution does occur, without paying attention to the specifics of the process, and to follow the ensuing chemical evolution\footnote{Note that our transport boundary condition necessarily leads us to overestimate the thickness of the pure-hydrogen layer at the surface. By definition, our float-up simulations can only produce hydrogen-layer masses larger than 10$^{-14} M$ (see for instance Figure \ref{fig:hprof_060_hot}), while true hydrogen-layer masses are likely smaller than that. In fact, it is thought that only hydrogen-layer masses smaller than 10$^{-14} M$ give rise to convective dilution \citep{rolland2018}, so our only way to study the outcome of this process is to assume that it takes place and to induce it artificially. This should not be interpreted as a physical statement that convective dilution occurs for hydrogen-layer masses larger than 10$^{-14} M$, but rather as a numerical procedure designed to overcome the shortcoming related to our transport boundary condition. Besides, note that this shortcoming does not affect the predicted hydrogen abundance following convective dilution, because the pure-hydrogen layer represents a very small fraction of the total amount of hydrogen in the star (see text).}. To achieve this in practice, we take the last stellar model of a given float-up simulation and artificially dilute its hydrogen-rich layer within a small helium-rich region underneath ($-14.0 < \log q < -12.5$). We then feed this modified model back into our code and resume the evolutionary calculation at $\Teff \sim 32,000$ K, just as the helium convection zone appears (see Figure \ref{fig:convzone}). In this regime, element transport is entirely dominated by the rapidly expanding convection zone, so we turn off atomic diffusion and mass loss for computational simplicity. We apply this procedure only to our most hydrogen-deficient simulations with $\log X_{\rm H,0} \le -5.0$, for which the superficial hydrogen layer remains thin and thus the assumption of convective dilution is safe. We want to stress that our crude description of the onset of the convective dilution mechanism does not affect the evolution of the hydrogen abundance profile at later times. This is because the bulk of the hydrogen is still located in the diffusion tail ($-10.0 \lta \log q \lta -4.0$; see for instance Figure \ref{fig:hprof_060_hot}), hence making the details of the outer stratification unimportant for our purpose. Furthermore, note that our method incorporates a number of significant improvements over that of \citet{rolland2018,rolland2020}: (1) we use realistic chemical profiles obtained from comprehensive calculations of the float-up process, (2) we rely on a full evolutionary approach in which white dwarf cooling and element transport are self-consistently coupled, and (3) we include mixing due to convective overshoot.

\begin{figure*}
\centering
\includegraphics[width=2.\columnwidth,clip=true,trim=1.75cm 8.25cm 1.75cm 10.00cm]{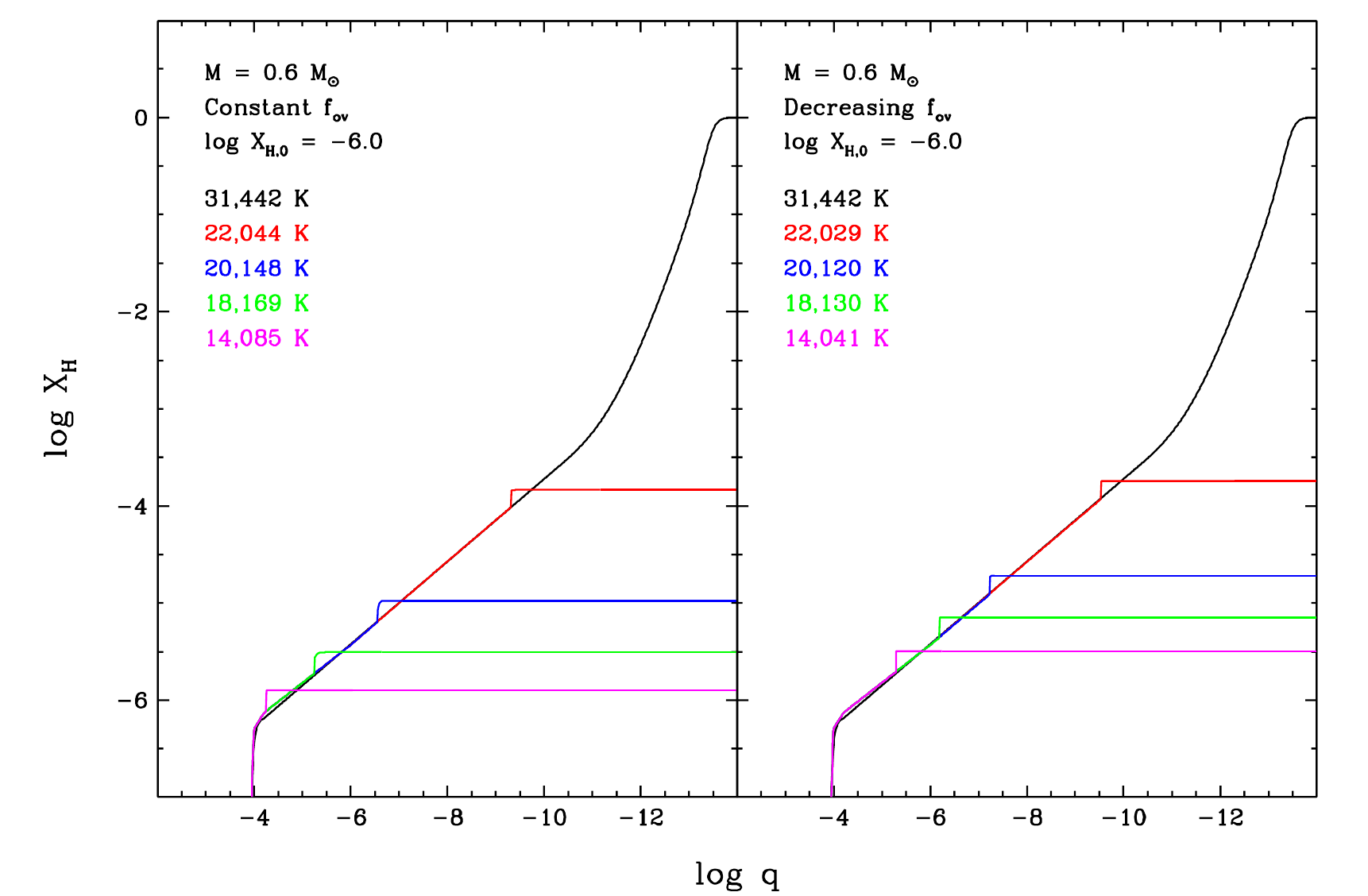}
\caption{Evolution of the hydrogen abundance profile following the convective dilution process in a white dwarf model with a mass $M = 0.6 \ \msun$, a constant (left panel) or decreasing (right panel) overshoot parameter (see Section \ref{sec:comp} for details), and an initial hydrogen abundance $\log X_{\rm H,0} = -6.0$. The hydrogen abundance is expressed as a mass fraction, while the position inside the star is measured in terms of the fractional mass depth ($q = 1-m/M$). Each curve represents a selected stage along the evolutionary sequence and is labeled with the corresponding value of the effective temperature.}
\vspace{2mm}
\label{fig:hprof_060_cool}
\end{figure*}

Figure \ref{fig:hprof_060_cool} displays the evolution of the hydrogen abundance profile at low effective temperature for the simulation started from the initial condition $\log X_{\rm H,0} = -6.0$. The left and right panels show the results obtained under the assumption of a constant and decreasing overshoot parameter, respectively, as described in Section \ref{sec:comp}. In both panels, the first curve represents the chemical structure prior to convective dilution (in this case, this corresponds to the last model displayed in the left panel of Figure \ref{fig:hprof_060_hot}), while the other four curves illustrate the chemical structure at $\Teff \sim 22,000$, 20,000, 18,000, and 14,000 K. Within the convection zone and the overshoot region, the convective motions mix hydrogen and helium very efficiently and thereby produce a flat hydrogen abundance profile. As the star cools and the convection zone grows inward, the hydrogen finds itself diluted within increasingly helium-rich layers, thus causing the surface hydrogen abundance to decrease. However, and this is a key point for what follows, the presence of the hydrogen reservoir in the deep envelope prevents the surface hydrogen abundance from dropping to exceedingly low values. Indeed, it is clear from Figure \ref{fig:hprof_060_cool} that the amount of hydrogen in the atmosphere is essentially set by the amount of hydrogen at the base of the convectively mixed region, which is non-zero given the broad hydrogen diffusion tail shaped by the float-up process. On another note, it is also apparent that the treatment of convective overshoot has a substantial impact on the atmospheric composition below $\Teff \sim 20,000$ K. More specifically, if the overshoot parameter is assumed to decrease with cooling, the homogeneously mixed region is less extended and therefore the surface hydrogen abundance is higher (by almost 0.5 dex), as expected from Figure \ref{fig:convzone}.

\begin{figure*}
\centering
\includegraphics[width=2.\columnwidth,clip=true,trim=1.75cm 6.75cm 1.75cm 8.75cm]{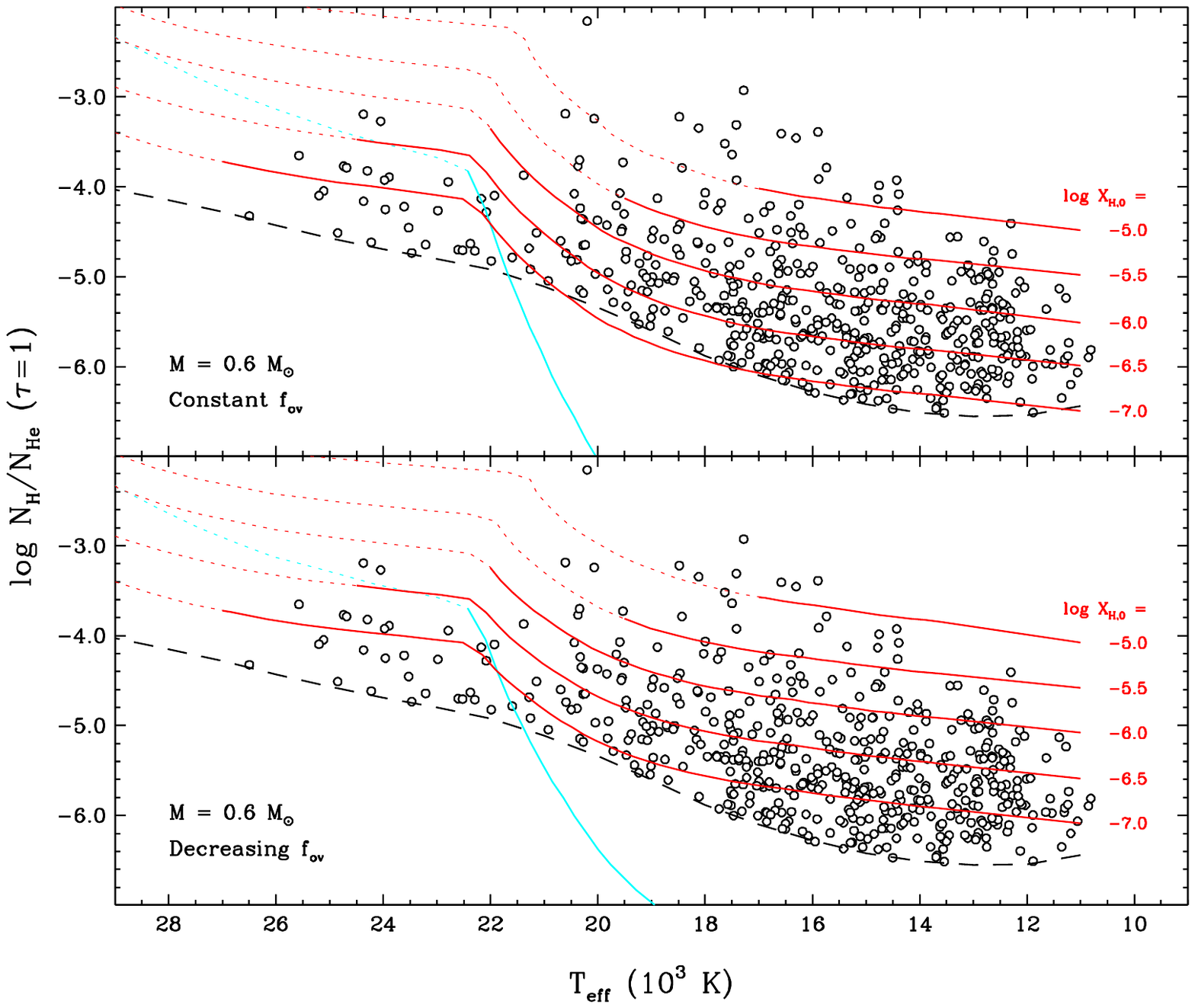}
\caption{Atmospheric hydrogen abundance as a function of effective temperature following the convective dilution process in white dwarf models with a mass $M = 0.6 \ \msun$, a constant (top panel) or decreasing (bottom panel) overshoot parameter (see Section \ref{sec:comp} for details), and various initial hydrogen abundances given in the figure (red curves). The instantaneous hydrogen abundance (the vertical axis) is measured in terms of the hydrogen-to-helium number ratio, while the initial hydrogen abundance (the label of each curve) is expressed as a mass fraction. For comparison, we display a second sequence with $\log X_{\rm H,0} = -6.0$ in which the deep hydrogen reservoir was artificially removed (cyan curves). The dotted parts of the curves indicate the region where the results are unreliable due to our artificial initiation of convective dilution. Also shown are empirical atmospheric parameters of DBA white dwarfs (\citealt{rolland2018,genest-beaulieu2019}; black circles) and the optical detection limit of hydrogen in a helium-rich atmosphere (dashed black line).}
\vspace{2mm}
\label{fig:evol_cool}
\end{figure*}

Figure \ref{fig:evol_cool} displays the surface hydrogen abundance as a function of effective temperature following convective dilution in our most hydrogen-deficient models. The top and bottom panels correspond to the cases of a constant and decreasing overshoot parameter, respectively. The dotted parts of the curves indicate the region where the results are unreliable due to our artificial initiation of convective dilution; modeling this region correctly would require a better understanding of the details of the mechanism. Nevertheless, for the reasons given above, the solid parts of the curves are physically accurate. In both panels, we also display as circles the DBA white dwarfs with empirically derived atmospheric parameters from \citet{rolland2018} and \citet{genest-beaulieu2019}. We express here the hydrogen abundance as the hydrogen-to-helium number ratio $N_{\rm H}/N_{\rm He}$, which is the usual observational quantity. This figure shows that regardless of the large uncertainty associated with the overshoot prescription, our simulations nicely reproduce the range of atmospheric compositions measured in cool DBA stars. This is in stark contrast with traditional models of the convective dilution process, which predict much lower surface hydrogen abundances \citep{macdonald1991,rolland2018,rolland2020}. This is illustrated in Figure \ref{fig:evol_cool} by the cyan lines, which are evolutionary sequences with $\log X_{\rm H,0} = -6.0$ in which we artificially removed the deep hydrogen reservoir before convective dilution, thereby replicating the behavior of these older models.

The crucial feature of the present calculations is the use of accurate chemical profiles obtained from a detailed assessment of the float-up process. Indeed, the problem of traditional models lies with their assumption that the thin pure-hydrogen layer that undergoes convective dilution represents all of the hydrogen contained within the white dwarf. In other words, they presume that gravitational settling is so efficient that a complete separation of hydrogen and helium has been previously achieved, such that that there remains no hydrogen below the superficial layer. Consequently, when the thin pure-hydrogen shell is diluted within the much more massive pure-helium envelope, the surface hydrogen abundance becomes extremely low. It turns out, as first suggested by \citet{rolland2020} and as convincingly demonstrated here, that this assumption is far from accurate: there is actually a large amount of residual hydrogen ``hidden'' below the superficial layer. Upon convective dilution, this hydrogen reservoir finds itself mixed within the convection zone and thus determines the atmospheric composition. \citet{rolland2020} refer to this phenomenon as a dredge-up process, but we have realized that this expression is somewhat of a misnomer, given that the concept of dredge-up generally involves an increase of the surface abundance of some trace element with time. As illustrated in Figure \ref{fig:hprof_060_cool}, we really are dealing with a dilution process here; the novelty of the current paradigm is simply that the superficial hydrogen layer is diluted within a helium-rich but hydrogen-contaminated (rather than hydrogen-free) envelope. For this reason, we believe that convective dilution remains an appropriate term to describe the phenomenon. Finally, we also note that the comparison between the predicted and observed hydrogen abundances shown in Figure \ref{fig:evol_cool} is much more satisfactory than that shown in Figure 11 of \citet{rolland2020}, a result that can be ascribed to our improved chemical profiles. Based on this excellent agreement, we can confidently state, for the first time, that the convective dilution scenario does provide a valid explanation for the atmospheric composition of the bulk of DBA white dwarfs.

\begin{figure*}
\centering
\includegraphics[width=2.\columnwidth,clip=true,trim=1.75cm 6.75cm 1.75cm 8.75cm]{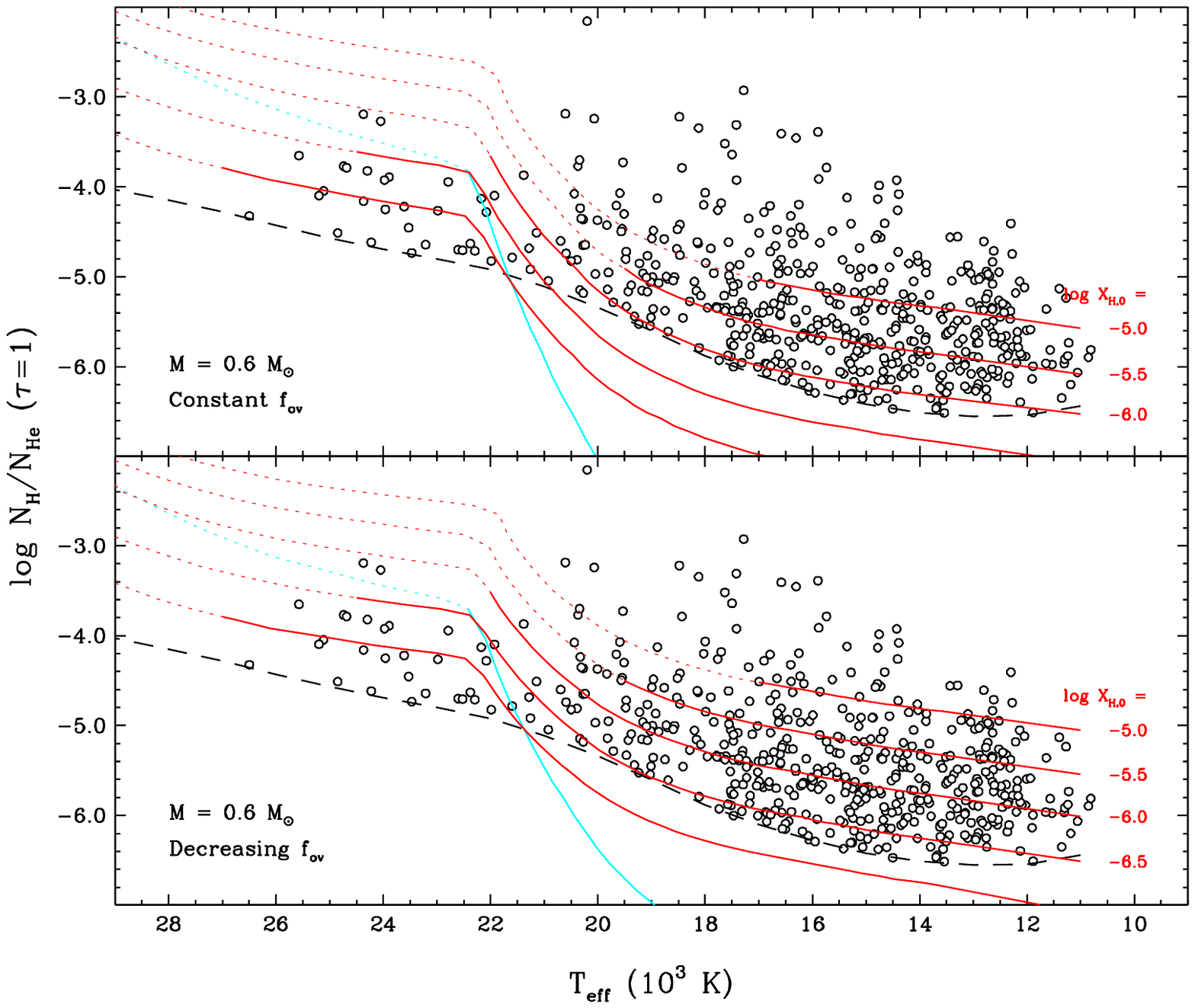}
\caption{Same as Figure \ref{fig:evol_cool}, but for models in which the fractional mass depth of the initial hydrogen reservoir (and thus the total mass of hydrogen) was arbitrarily reduced by a factor of 10.}
\vspace{2mm}
\label{fig:evol_cool_2}
\end{figure*}

As mentioned in Section \ref{sec:comp} and illustrated in Figure \ref{fig:hprof_060_cool}, the predicted hydrogen abundance following convective dilution should depend on the assumed depth of the hydrogen reservoir at the very beginning of our float-up simulations. In all of the evolutionary sequences presented so far, the residual hydrogen is initially mixed down to $\log q = -4.0$. This is roughly the largest possible depth (given that deeper-lying hydrogen is expected to be burned) and thus translates into the most massive possible reservoir for a given initial abundance. This choice effectively maximizes the atmospheric hydrogen content in the DBA phase, and it is legitimate to ask whether the conclusion of the present section still holds if the initial hydrogen reservoir is shallower. To answer this question, we repeated the calculations shown in Figure \ref{fig:evol_cool} but starting from models in which trace hydrogen is included only above $\log q = -5.0$. This amounts to reducing the total mass of hydrogen by a factor of 10 with respect to our standard case, which is a rather extreme assumption. The results of this experiment are displayed in Figure \ref{fig:evol_cool_2}, which is otherwise identical to Figure \ref{fig:evol_cool}. As expected, the theoretical curves are shifted downward, most notably at low temperature, when the convection zone reaches the deep envelope. Yet, despite our significant reduction of the size of the hydrogen reservoir, our simulations still reproduce the observed atmospheric composition of most DBA stars, especially with the decreasing-$f_{\rm ov}$ prescription. This confirms the robustness of our qualitative statement that the majority of DBA white dwarfs can be interpreted as products of the convective dilution mechanism. However, Figures \ref{fig:evol_cool} and \ref{fig:evol_cool_2} show that a quantitative analysis of individual DBA stars is not possible at the present time given the uncertainties on the extent of the overshoot region and of the hydrogen reservoir.

\section{Discussion} \label{sec:disc}

\subsection{The Hydrogen Content of Hot DO White Dwarfs} \label{sec:disc_DO}

In Paper I, we performed a spectroscopic analysis of a large sample of hot white dwarfs in order to study the spectral evolution of these objects from an empirical perspective. We were able to determine that (1) about one in four stars enter the cooling sequence with a helium-rich atmosphere, and that (2) about two-thirds of them develop a hydrogen-rich atmosphere by the time they reach $\Teff \sim 30,000$ K. Moreover, we found that different objects experience this DO/DB-to-DA transition at different effective temperatures over the range $70,000 \ {\rm K} \gta \Teff \gta 30,000$ K. If we interpret the atmospheric transformation as the consequence of the float-up of residual hydrogen, the results of Paper I suggest that the population of hot helium-dominated white dwarfs is characterized by a broad range of hydrogen content. But what is this range exactly? In other words, what is the quantitative relation between the amount of residual hydrogen and the transition temperature? It has been the primary goal of the present work to answer this question through theoretical calculations of element transport in evolving white dwarfs.

Our simulations of the float-up process have indeed provided a completely new way to estimate the hydrogen abundance left in the stellar envelope at the beginning of the cooling sequence, although this inference depends on the rather uncertain treatment of the radiative wind that opposes gravitational settling. As shown in Section \ref{sec:res_float}, the range of initial hydrogen abundances that best reproduces the observed spectral evolution is $-6.0 \lta \log X_{\rm H,0} \lta -3.0$ for our strong-wind model and $-7.0 \lta \log X_{\rm H,0} \lta -4.0$ for our weak-wind model (which are roughly representative of objects with PG 1159-type and O(He)-type progenitors, respectively). As a corollary, the upper limit on the initial hydrogen abundance of those few white dwarfs that always retain a helium-rich surface is $\log X_{\rm H,0} \lta -6.0$ and $-7.0$ for the two mass-loss prescriptions. 

Although these constraints are of great significance (as further discussed below), the ultimate quantity of interest is the total mass of hydrogen present in the star, which unfortunately is not straightforward to obtain. The first and foremost reason for this is that the conversion from an abundance to a total mass hinges on our assumption regarding the depth to which hydrogen is initially mixed, as noted in Section \ref{sec:comp}. In our simulations, we adopt the maximum possible depth, $\log q = -4.0$, which means that the total mass of hydrogen is simply given by $M_{\rm H}/M = 10^{-4} X_{\rm H,0}$. Consequently, our models undergoing the DO/DB-to-DA transition in the range $70,000 \ {\rm K} \gta \Teff \gta 30,000$ K are characterized by $-10.0 \lta \log M_{\rm H}/M \lta -7.0$ in the strong-wind case and $-11.0 \lta \log M_{\rm H}/M \lta -8.0$ in the weak-wind case. Nevertheless, we cannot exclude the possibility that the residual hydrogen is mixed less deeply; for instance, according to Figure 3 of \citet{althaus2005b}, the base of the hydrogen reservoir may be located up to $\sim$0.5 dex higher in $\log q$, which corresponds to a $\sim$0.5 dex uncertainty in $\log M_{\rm H}/M$. Furthermore, another issue is that the total amount of hydrogen is expected to decrease with time as a result of nuclear burning at the bottom of the envelope and mass loss at the surface. In our calculations, the first effect is ignored, while the second effect is highly uncertain. To illustrate the latter point, we draw attention to the float-up simulation presented in Section 3.2 of Paper II. In that work, we stated that as much as $\sim$90\% of the hydrogen can be ejected by the wind in the course of the evolution. Upon further analysis, we have realized that most of this mass loss occurs at relatively low effective temperature ($\Teff \lta 40,000$ K) due to the fact that the decrease in mass-loss rate is then compensated by the increase in evolutionary timescale\footnote{As a rough example, for a typical model at $\Teff \sim 30,000$ K, the mass-loss rate obtained from Equation \ref{eq:mdot} is $\sim$$5 \times 10^{-17} \msun \rm{yr}^{-1}$ and the evolutionary timescale is $\sim$$1 \times 10^{7} \rm{yr}$, and thus the mass of hydrogen lost in a single timestep may be as large as $\sim$$5 \times 10^{-10} \msun$, which may represent a significant fraction of the total amount of hydrogen in the star.}. In reality, the radiative wind is expected to die out quite quickly with cooling and thus should not have any influence at such a late time, so this particular result is doubtful and likely stems from our extrapolation of the mass-loss law well beyond its domain of validity. Although the calculations of the present paper do not suffer from this problem (given that, as mentioned in Section \ref{sec:res_conv}, we turn off the wind at low effective temperature), the fact remains that our treatment of the wind is very rudimentary. For all of these reasons, we refrain from further interpreting our results in terms of total masses of hydrogen.

Despite this limitation, our estimate of the hydrogen abundance in hot helium-rich white dwarfs is still noteworthy as it provides independent information on the phenomena responsible for the hydrogen deficiency in the first place. Two main mechanisms are believed to produce hydrogen-poor pre-white dwarf objects: the occurrence of a late helium-shell flash in a post-AGB star, which then becomes a PG 1159 star \citep{iben1983,herwig1999,althaus2005a,werner2006}, and the merger of two low-mass white dwarfs, which possibly leads to the formation of a O(He) star \citep{zhang2012,reindl2014a}. In both cases, it is practically impossible to detect low hydrogen abundances spectroscopically ($\log X_{\rm H,0} \lta -2.0$), because all hydrogen lines are blended with ionized helium lines \citep{werner1996}. For this reason, the exact amount of residual hydrogen can only be assessed through theoretical models of the aforementioned processes, which are extremely challenging to compute and hence come with their share of uncertainties. We can compare our hydrogen abundance constraints, which have been derived from a totally different approach, with the predictions of such calculations. Models of the late helium-shell flash have shown that there exist two common variants of this phenomenon, which are generally referred to as the late thermal pulse (LTP) and very late thermal pulse (VLTP) scenarios (see for instance \citealt{werner2006}). The LTP version leads to relatively large hydrogen abundances ($\log X_{\rm H,0} \gta -2.0$), so it definitely cannot account for the DO white dwarfs that turn into DA white dwarfs in the range $70,000 \ {\rm K} \gta \Teff \gta 30,000$ K\footnote{We do not imply here that the LTP scenario does not occur in nature, but rather that it results in a hydrogen abundance large enough that the star likely develops a hydrogen-rich atmosphere before even reaching the white dwarf cooling sequence. The so-called hybrid PG 1159 stars, in which hydrogen is spectroscopically visible, are extreme examples of such objects.}. On the other hand, these objects appear to be well explained by the VLTP version. For instance, according to the VLTP models of \citet{miller-bertolami2006}, the hydrogen abundance immediately after the helium-shell flash is $-5.3 \lta \log X_{\rm H,0} \lta -3.4$ depending on the adopted constitutive physics (see their Table 2). Our results are entirely consistent with these values, but additionally indicate that the hydrogen content left intact by the VLTP event varies from star to star. The preliminary calculations of \citet{miller-bertolami2017} suggest that this could be the consequence of differences in main-sequence mass and/or metallicity. Besides, we note that most theoretical studies of the VLTP scenario have claimed that the leftover hydrogen is probably entirely shed off by mass loss in the giant phase before the star returns to the cooling sequence \citep{iben1983,althaus2005a,werner2006}; obviously, the spectral evolution of hot white dwarfs strongly challenges this assertion. Finally, as for the channel involving the merger of two low-mass white dwarfs, the available models, such as those of \citet{zhang2012}, usually assume that all of the hydrogen is destroyed and therefore make no prediction regarding the remaining hydrogen content.

\subsection{The Origin of Hydrogen in Cool DBA White Dwarfs} \label{sec:disc_DBA}

There is a general consensus that the increase in the fraction of helium-rich white dwarfs at low effective temperature is due to the convective dilution and convective mixing processes. On the other hand, the origin of the trace hydrogen seen at the surface of the majority of these objects has been a matter of debate for a few decades. There currently exist two main schools of thought regarding the atmospheric composition of DBA stars: either the hydrogen is primordial, or it has been accreted. The first scenario, in which the surface composition is simply the outcome of convective dilution, potentially has the merit of explaining both the spectral evolution and the DBA phenomenon with a single mechanism. However, the hydrogen abundances predicted by traditional models of the convective dilution process are notoriously much too low \citep{macdonald1991,rolland2018,rolland2020}. For this reason, most recent works on DBA white dwarfs have favored the second scenario, more specifically the accretion of water-rich planetesimals \citep{raddi2015,gentile-fusillo2017,cunningham2020}. Meanwhile, \citet{rolland2020} suggested that the apparent failure of the primordial hypothesis might only be an unfortunate consequence of the inaccuracy of previous studies of convective dilution. They argued that the atmospheric hydrogen of DBA stars may actually come from an internal reservoir of residual hydrogen, which is expected from the float-up process but had been completely overlooked in the past.

We have investigated this idea using our comprehensive simulations of the float-up process at high effective temperature, which constitute a significant improvement with respect to the preliminary models of \citet{rolland2020}. Our calculations indeed predict the existence of a massive hydrogen reservoir underneath the thin pure-hydrogen layer. As demonstrated in Section \ref{sec:res_conv}, this key feature of the chemical structure effectively resolves the issue of the primordial scenario: following convective dilution, our models closely match the range of surface hydrogen abundances observed among the DBA population. These results compellingly validate the paradigm put forward by \citet{rolland2020}. In a nutshell, the convective dilution mechanism does produce the atmospheric composition of DBA white dwarfs, provided that realistic hydrogen abundance profiles are considered. Furthermore, we want to stress that this conclusion is independent of the details of how the superficial pure-hydrogen layer gets mixed within the underlying helium-rich envelope, because the final surface hydrogen abundance is largely governed by the deep hydrogen reservoir. However, this also implies that the predicted abundances are subject to the uncertainties on the exact size of this hydrogen reservoir (related to initial conditions, nuclear burning, and mass loss, as discussed in the previous subsection), in addition to the uncertainty associated with convective overshoot. In other words, the quantitative relation between the initial and final surface hydrogen abundances of a given simulation should be taken with great caution. Nevertheless, this shortcoming does not affect the qualitative conclusion that our models provide a natural explanation for the atmospheric composition of DBA white dwarfs.

In light of this development, it is worth reviewing the strengths and weaknesses of each of the two theories commonly invoked in this context. In principle, the accretion of water-rich planetesimals remains a viable origin for the hydrogen of DBA stars. There is even undeniable observational evidence that this phenomenon indeed occurs in a few objects (\citealt{gentile-fusillo2017} and references therein). The question to be asked is whether water accretion is responsible for the entire DBA population. As mentioned earlier, one drawback of this scenario is the requirement that hydrogen be delivered to the star only after the convective dilution process has taken place. However, this difficulty is not insurmountable: it has recently been pointed out by \citet{hoskin2020} that the dynamical accretion simulations of \citet{mustill2018} predict that planetesimals cannot be scattered to within the Roche limit of a white dwarf for the first $\sim$40 Myr of its cooling lifetime, which corresponds to $\Teff \gta 22,000$ K according to our evolutionary calculations. This delay in the onset of accretion is exactly what is needed to produce cool DBA white dwarfs. Nevertheless, another complication is that DBA stars represent as much as 60$-$75\% of the DB population \citep{koester2015,rolland2018}. Within the accretion hypothesis, this means that 60$-$75\% of all white dwarfs must accrete water-bearing planetesimals in their first few 100 Myr on the cooling sequence (assuming that the phenomenon is equally common in hydrogen-rich and helium-rich objects), a number which is uncomfortably high. Still, this scenario has generally been favored until now due to the lack of a better alternative, that is, because of the failure of the model based on residual hydrogen to account for the measured abundances of DBA stars.

Now that this problem has been fully resolved, perhaps it is time to question the relevance of the water-accretion interpretation for the bulk of the DBA population. Our calculations have demonstrated that DBA white dwarfs actually represent an inevitable evolutionary stage of stars retaining a small amount of primordial hydrogen on the cooling sequence. From this new perspective, the chief advantage of the primordial scenario is that it consistently explains (1) the helium-to-hydrogen transformation at high effective temperature, (2) the hydrogen-to-helium transformation at low effective temperature, and (3) the atmospheric composition of DBA white dwarfs. In contrast, the accretion scenario explains only the last item\footnote{One could allege that hydrogen accretion could possibly explain the DO/DB-to-DA transition at high effective temperature, but this would contradict the argument according to which a significant delay in the onset of accretion is necessary to explain the existence of DBA white dwarfs.}. Furthermore, the results of Paper I indicate that about two-thirds of hot DO stars contain enough residual hydrogen to transform into DA white dwarfs and then into DBA white dwarfs; this is remarkably consistent with the observed proportion of DBA stars among the DB population quoted in the previous paragraph. Therefore, although the accretion of water-rich planetesimals undoubtedly plays a role in some individual cases (especially in the most hydrogen-rich objects with $\log N_{\rm H}/N_{\rm He} \gta -4.0$), it now seems somewhat superfluous to invoke this additional mechanism as the primary source of hydrogen. In short, our study strongly supports the idea that the trace hydrogen seen at the surface of DBA white dwarfs is predominantly of primordial origin.

In recent years, a further argument in favor of widespread water accretion onto DBA stars has often been put forward. It was shown by \citet{gentile-fusillo2017} that there exists a clear correlation between the presence of hydrogen and the presence of metals in helium-dominated atmospheres. Given that the heavy elements are necessarily accreted, they interpreted this correlation as evidence that the hydrogen is accreted alongside the metals, most plausibly in the form of water. The obvious question then is whether this fact can be reconciled with our above conclusion. We believe so, as there is another possible explanation for the observed correlation, the source of which lies in the pre-white dwarf evolution. Stating that the presence of hydrogen and the presence of metals are related is equivalent to stating that the absence of hydrogen and the absence of metals are related. In other words, those DB white dwarfs that appear devoid of hydrogen tend to appear devoid of planetary material as well; let us call them the pure DB stars. Given that the very existence of hydrogen-deficient white dwarfs in general can be traced back to some violent event before the cooling phase, either a late helium-shell flash or a stellar merger, it is reasonable to think that the extreme hydrogen deficiency of the pure DB stars is simply the result of an extreme version of these phenomena. For instance, a particularly violent helium-shell flash may deplete the primordial hydrogen content of the star to such a low level that it will never be detectable at any point on the cooling sequence. Interestingly enough, another potential consequence of a more active evolutionary history is that a planetary system may be less likely to survive until the white dwarf phase. In the example of the late helium-shell flash, it is not difficult to imagine that the rapid episodes of expansion and contraction of the pre-white dwarf object may lead to the ejection or engulfment of any small orbiting body and thus prevent the accretion of planetary material at later times. The conclusion is that we indeed expect the pure DB white dwarfs to be strongly deficient in both hydrogen and metals, regardless of the composition of their former planetary companions. This, of course, naturally gives rise to the observed trend that DBA stars often exhibit traces of heavy elements, which therefore cannot be taken as a definitive proof of water accretion onto white dwarfs. That said, we recognize that the scenario outlined here is somewhat speculative and will have to be confirmed by appropriate orbital dynamics simulations.

\section{Summary and Conclusion} \label{sec:conclu}

In this paper, we presented detailed theoretical calculations of the transport of residual hydrogen in helium-dominated white dwarfs. We first studied the upward diffusion of trace hydrogen at high effective temperature, which eventually leads to the formation of a hydrogen-rich atmosphere. We examined how this float-up mechanism is influenced by the initial hydrogen abundance and the strength of the radiative wind that competes with gravitational settling. We showed that the atmospheric transformation takes place later in the cooling process when a lower hydrogen content or a higher wind mass-loss rate is assumed. Based on these results, we interpreted the empirical fact that DO/DB stars turn into DA stars over a broad range of effective temperatures as a consequence of the existence of a broad range of hydrogen content among the hot helium-rich white dwarf population. More specifically, we found that the initial hydrogen abundances that best explain the observed spectral evolution are $-6.5 \pm 0.5 \lta \log X_{\rm H,0} \lta -3.5 \pm 0.5$ if we allow for uncertainties in the strength of the wind. These values are roughly consistent with the level of hydrogen deficiency expected from models of late helium-shell flashes in post-AGB stars. Furthermore, we noted that our simulations predict that the thin hydrogen shell at the surface represents only a small fraction of the total amount of residual hydrogen, as most of it remains ``hidden'' in the deep envelope for a long period of time.

We then investigated the convective dilution of the superficial hydrogen layer within the underlying helium-dominated envelope at low effective temperature. We demonstrated that our models successfully reproduce the range of atmospheric hydrogen abundances observed among cool DBA white dwarfs, in stark contrast with previous studies of convective dilution. This remarkable improvement is essentially due to the use of realistic chemical profiles obtained from the float-up process, including in particular the deep hydrogen reservoir which had been completely overlooked until the recent work of \citet{rolland2020}. By establishing that the existence of DBA stars is an inevitable consequence of spectral evolution, our calculations provide a natural solution to the three-decade-old problem of the origin of hydrogen in these objects, arguably the most challenging problem of spectral evolution theory. This outcome is perhaps the most striking example of the importance of considering a self-consistent, time-dependent treatment of element transport in evolutionary models of white dwarfs. We thus concluded that the atmospheric composition of the majority of DBA stars can be explained solely by the convective dilution of primordial hydrogen, without the need to invoke external accretion. Nevertheless, we noted that the quantitative relation between the initial and final surface hydrogen abundances of a given object is still uncertain, especially because it depends on the location of the base of the hydrogen reservoir and on the efficiency of convective overshoot. Better constraints on these two physical ingredients, as well as on the strength of the wind at high temperature, would definitely help us refine our understanding of the spectral evolution of white dwarfs.

\acknowledgments

We dedicate this paper to the memory of our late colleague, mentor, and friend, Gilles Fontaine, who would certainly have been very enthusiastic about the results presented here. We thank the anonymous referee for a thorough and constructive report that helped improve the quality of the paper. This work was supported by the Natural Sciences and Engineering Research Council (NSERC) of Canada and the Fonds de Recherche du Qu\'ebec $-$ Nature et Technologie (FRQNT).

\bibliographystyle{aasjournal}
\bibliography{main}

\end{document}